\documentstyle[11pt,newpasp,twoside,epsf]{article}
\def\plotone#1{\centering \leavevmode\epsfxsize=\columnwidth \epsfbox{#1}}

\begin{document}

\title{Current Development of the Photo-Ionization Code Cloudy}
\author{P.A.M. van Hoof$^{1,2}$, P.G. Martin$^1$, G.J. Ferland$^2$}
\affil{$^1$Canadian Institute for Theoretical Astrophysics,
University of Toronto, McLennan Labs, 60 St. George Street,
Toronto, ON M5S 3H8, Canada}
\affil{$^2$University of Kentucky, Department of Physics \& Astronomy, 177 CP
Building, Lexington, KY 40506--0055, USA}

\begin{abstract}
This paper describes the development of the photo-ionization code Cloudy
since the last major release 90.05.
\end{abstract}

\section{Introduction}

This poster centers on the development of Cloudy, a large-scale code designed
to compute the spectrum of gas in photo-ionization or collisional balance.
Such plasma is far from equilibrium, and its conditions are set by the balance
of a host of micro-physical processes. The development of Cloudy is a
three-pronged effort requiring advances in the underlying atomic data base,
the numerical and computational methods used in the simulation, and
culminating in the application to astronomical problems. These three steps are
strongly interwoven.

In recent years, Cloudy has undergone a major upgrade in the atomic data. A
large part of this effort was concentrated on improving the photo-ionization
cross sections. Since then the code has been transformed from a Fortran 77 to
an ANSI C code and work has begun to completely recode the model atoms used to
calculate line emissivities. They will be organized in such a way that each
iso-electronic sequence has a common code appropriate for all members of that
sequence. This makes validating the code easier and also facilitates future
upgrades in the atomic data. Work on the hydrogen sequence is complete and
work on the helium sequence is underway. A more detailed account of the
current status of the code can be found in Section~2. The grain model in
Cloudy is also currently undergoing a major upgrade, which is described in
more detail in Section~3. 

\section{Improvements in Cloudy 94}

Cloudy has undergone a major transformation in the past year with respect to
the last release Cloudy~90. The most important differences are listed below. A
first release of Cloudy~94 is currently available from the Cloudy web-site (for
details see Section~4).

\begin{itemize}
\item
  The code is now strict ANSI/ISO 89 C. As a result Cloudy is now
    exceptionally GNU gcc and Linux friendly. After this version is released
    the development version will move to C++ as gcc evolves onto the ANSI/ISO
    C++ standard and the C++ standard template library matures.
\item
  All hydrogenic species H\,{\sc i} through Zn\,{\sc xxx}, and their
    respective ions are treated with a common model atom that uses a single
    code base. This atom reproduces accurate hydrogenic emissivity to within
    the uncertainties in the atomic data. The 30 hydrogenic ions can have up
    to 400 levels.
\item
  The continuum now extends down to 10$^{-8}$~Ryd ($\lambda\approx 10$~m).
    This is needed because the continuum must extend to the energy of the 400
    -- 399 transition of hydrogen.
\item
  Previous versions used simple approximations for the hydrogenic ionization
    and level balance for temperatures too low to compute NLTE departure
    coefficients. The new version determines level populations for low
    temperatures rather than departure coefficients, so low temperature
    predictions are as valid as high temperature results.
\item
  Two additional sets of NLTE stellar atmospheres are available -- the CoStar
    grid of wind-blanketed O-stars, and the Rauch low-metallicity grid of
    white dwarf atmospheres.
\item
  The optimization algorithm PHYMIR is included. This allows optimization runs
    to be executed much more efficiently on parallel UNIX computers by
    calculating individual models simultaneously on different processors. This
    can drastically reduce the amount of wallclock time needed, depending on
    the number of free parameters.
\item
  The ionization/thermal kernel has been totally rewritten to incorporate all
    the lessons learned from known convergence problems. As a result C94 is
    more stable than C90, with far better convergence properties.
\item
  All previous versions only considered ionization stages that could be
    produced by the incident continuum. This limit has been lifted, so
    collisional or coronal equilibrium models can be computed with very soft
    incident continua.
\item
  Much of the code is now double precision. As a result the code will work for
    a broader range of densities than before. Densities well below
    10$^{-5}$~cm$^{-3}$ or above 10$^{17}$~cm$^{-3}$ can be computed without
    under/over flow.
\item
  The assert command has been introduced. This tells the code to verify that
    its predictions agree (within a stated uncertainty) with a known result.
    The test cases make extensive use of this feature, which provides an
    automatic way to validate the code. As of now, the entire test suite of
    standard models is recomputed and verified every single night.
\item
  All large storage items are dynamically allocated at run time, taking only
    the needed memory. As a result, in its default state, C94 actually takes
    less memory than C90. It also executes slightly faster than C90.
\end{itemize}

\section{The Grain Model}

\begin{figure}
\plotone{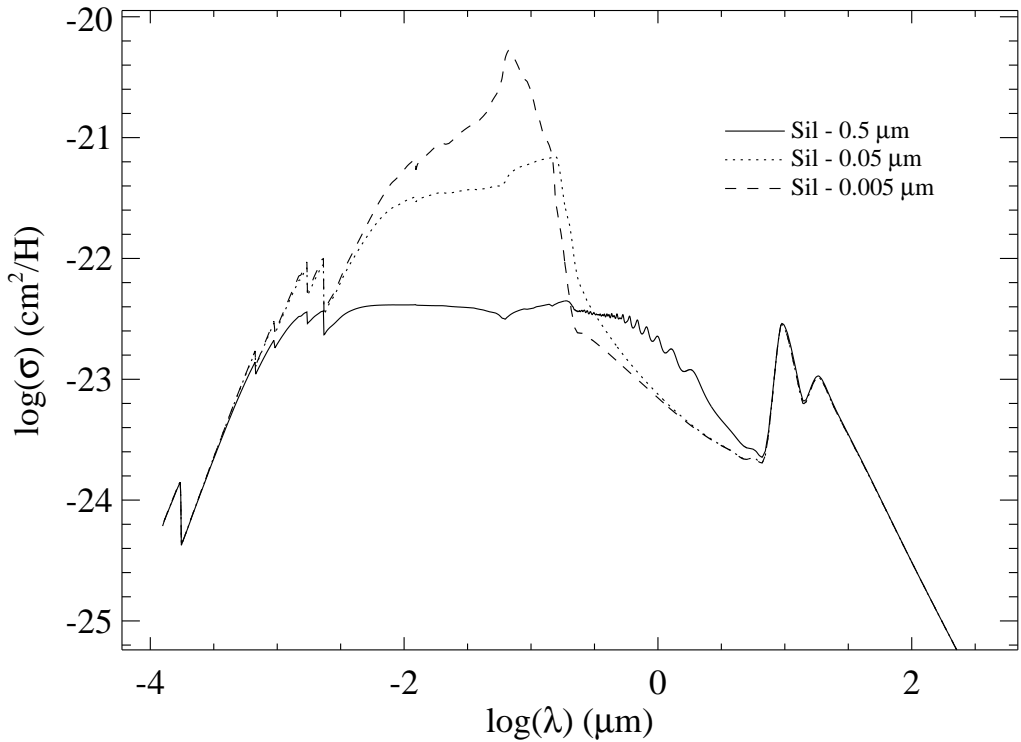}
\caption{ 
The absorption cross section for astronomical silicate for three
single sized grains. The dust-to-gas ratio is the same for all three species
and the cross sections are normalized per hydrogen nucleus in the plasma.}
\end{figure}

The current grain model was introduced to Cloudy in 1990 to facilitate more
accurate modeling of the Orion nebula (for a detailed description see
Baldwin et al., 1991, ApJ 374, 580). The model has since undergone some minor
revisions but remained largely the same. Recently, our knowledge of grains
has been greatly advanced by the results from the ISO mission. In view of
these rapid developments we have undertaken a major upgrade of the grain
model in Cloudy. The two main aims are to make the code more flexible and
to make the modeling results more realistic.

In particular, in the current model the grain opacities for a handful of grain
species are hard-wired in the code. Furthermore, only a single equilibrium
temperature is calculated for each species. This is inappropriate for a grain
size distribution since small grains will be hotter than large grains. This is
caused by the fact that grain opacities depend quite strongly on grain size,
as shown in Figure~1. To improve the model, the following changes are being
implemented:

\begin{itemize}
\item
We will include a spherical Mie code in Cloudy. The necessary
optical constants needed to run the code will be read from a separate file.
This allows greater freedom in the choice of grain species. Files with
optical constants for a range of materials will be included in the Cloudy
distribution. However, the user can also supply his own optical constants for
a completely new grain type.
\item
We will introduce mixing laws to the code. This will allow the user to
define grains which are mixtures of different materials. Cloudy will then
calculate the appropriate opacities by combining the optical constants of
these grain types. This will allow the user to simulate aggregate or
`fluffy' grains.
\item
The absorption and scattering opacities will be calculated by Cloudy for more
arbitrary grain size distributions. This will give the user considerably more
freedom. Currently only a standard ISM and a truncated Orion size
distribution are included in the code.
\item
The size distribution will be split up in many small bins, and an equilibrium
temperature will be calculated for each bin separately. This allows
non-equilibrium heating to be treated and more realistic grain emission
spectra to be calculated. First tests show that under realistic conditions
this can make large differences in the flux (at least a factor of two) in the
Wien tail of the grain emission. The total flux emitted by the grains remains
virtually unchanged however.
\end{itemize}

\section{Cloudy on the Web}

The source for Cloudy can be obtained from the web at the following URL:

\begin{center}
http://www.pa.uky.edu/{$\sim$}gary/cloudy
\end{center}

This site contains the current major release (Cloudy 90.05) as well as the
beta2 release of Cloudy 94 (actually called Cloudy 93.03).

You can add your name to the Cloudy mailing list by logging on to the
following URL:

\begin{center}
http://nimbus.pa.uky.edu/cloudy/versions.htm
\end{center}

\end{document}